# TOWARD VIRTUAL CERAMIC COMPOSITES

M. Genet[1], P. Ladevèze[1,2], G. Lubineau[1]
[1] LMT-Cachan (ENS-Cachan, CNRS, Paris 6 University, UniverSud Paris PRES)
61 avenue du Président Wilson – 94235 Cachan CEDEX
{genet, ladeveze, lubineau}@lmt.ens-cachan.fr
[2] EADS foundation chair, Advanced Computational Structural Mechanics

## SUMMARY

A first step toward a multi-scale and multi-physic model –a virtual material– for self-healing ceramic matrix composites is presented. Each mechanism –mechanical, chemical– that act on the material's lifetime at a given scale –fibre, yarn– is introduced in a single modeling framework, aimed at providing powerful prediction tools.

## INTRODUCTION

Snecma Propulsion Solide, a Safran group company, has developed a range of woven SiC/[Si-B-C] composites designated as self-healing materials [1]. The composites are build up from woven yarns of SiC fibers infiltrated by a multi-layered ceramic matrix. During the load, a first crack network appears in the inter-yarn matrix, perpendicularly to the principal traction [2] (see Figure 1). Once it is saturated, a second network, oriented by the fibers, appears in the intra-yarn matrix [2]. The fibers are mechanically protected through the use of a [C] interphase deviating the cracks [2, 3] –people talk about mechanical fuse–. This process lead to a great increase in the material's strength. However, the fibers might also suffer from sub-critical cracking under oxidizing atmosphere [4]. The self-healing process consists in filling the matrix cracks with the product of the oxidation of certain components of the matrix, limiting the diffusion of oxygen toward fibers [5] –people talk about chemical fuse–. This process leads to a great increase in the material's lifetime [5].

In order to reduce the cost of certification and optimization of materials and structures, the industry is in need for specific lifetime prediction tools: they must be extrapolating –valid for very large lifetimes, beyond experimental identification– and robust –valid for a whole family of materials with a light identification–.

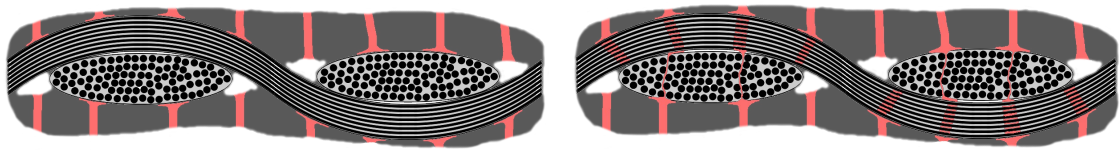

Figure 1: Crack network

Meso-models have been developed at Cachan in previous works [6,7,8,5]. An anisotropic damage macro-model with closure effect has been introduced [6] to predict the material's mechanical behavior at the scale of the structure with an *a priori* damage kinematics. Another model with non imposed damage kinematics has been developed [7] to facilitate the introduction of micro-informations. Then, a link from mechanical damage to density and opening of micro-cracks, as well as chemical mechanisms such as matrix self-healing and fiber degradation, have been introduced [8] to predict the material's lifetime. Developments and identification are still in progress [5]. However, the mechanical part of such models being completely macroscopic, they are limited in the description of the micro-mechanisms, although they play a major role in the material's lifetime. In this paper is presented a first version of a multi-scale and multi-physic model –called virtual material– for lifetime predictions on CMCs. The objective is to define a modelling framework large enough to introduce each mechanism – mechanical, chemical– that acts on the material's lifetime at a given scale –fibre, yarn– with a proper representation –continuum, discrete–. Strongly physics-based, this material database is expected to be suitable for extrapolation and robust. For now it has two main parts –mechanical and chemical–, which are presented in this paper in their actual state of development.

The mechanical part is presented section 2. It is based on an hybrid representation of the morphology of the crack network at yarn scale: explicit and homogenized representations are melted in a single framework so as to focus on essential informations. In addition to provide precise information on the mechanical state of the material along the load, this part provides essential data –mechanical load on fibers and oxygen path within the material– for the chemical models. The chemical part is presented section 3. For now, it consists in a model of the stress and oxidation induced sub-critical fracture of fibers, based on a framework coupling the mechanics and chemistry underlying the sub-critical propagation of cracks, as well as a model of fiber close environment. It now provides lifetime predictions for fibers, and should finally provide lifetime predictions for the material.

## MECHANICAL PART OF THE MODEL

In order to predict the mechanical state of the material and to derive input data for the chemical part, finite element calculations are run on meso-cells at yarn-scale. The finite element meshes, provided by LCTS [9], are very close to the reality of the woven yarns and the chemical vapor infiltrated matrix (see Figure 2). The following mechanisms are introduced within those cells, each one having a specific representation:

- Inter-yarn matrix cracks, being handled at their own scale and having a not *a priori* known orientations [2], are given an explicit representation using localization zones of a continuum anisotropic damage model. This model is presented section 2.1.

- Longitudinal intra-yarn matrix cracks, also being handled at their own scale – there are two or three cracks per yarn [10]– but having an *a priori* known orientations –along fibers– [2], are modeled explicitly using discrete crack surfaces [11, 12]. This has not been achieved yet, and will be presented later on.

- Transversal intra-yarn matrix cracks, being handled at a superior scale but having a substantial effect on both mechanical –through the debonding zone induced between fibers and matrix [13,3]– and chemical –through the arrival of oxygen on fibers [5]– micro-mechanisms, are modelled microscopically and homogenized through a complete micro-macro bridge [14], thus calculated macroscopically. This model is presented section 2.2.

**Matrix cracking**

Regarding the development of a damage model for the inter-yarn cracks, the following points have to be considered: i) the direction of the crack network is not known *a priori*: it is oriented by the external load; ii) when the material is in compression, it recovers its original stiffness as the cracks are closed.

Several attempts are discussed in the literature to deal with these problems [15]. The proposed model is build in the framework of the anisotropic damage theory [7]. Although a similar model has already been used as a complete macro-model –with inter- and intra-yarn matrix cracks– [6, 7, 8], it is used here for the inter-yarn matrix cracks only. The main idea is to split the elastic energy into a 'traction' energy and a 'compression' one:

$$e_d = e_t + e_c \quad \text{with} \quad \begin{cases} \dot{e}_t \\ \dot{e}_c \end{cases}$$

Where $\underline{\underline{C_0}}$ is the initial compliance operator and $\underline{\underline{C}}$ the damaged compliance operator for traction –it is an internal variable of the model, initially equal to $\underline{\underline{C_0}}$–. Moreover, $\underline{\underline{\sigma}}^+$ and $\underline{\underline{\sigma}}^-$ are the positive and negative parts of $\underline{\underline{\sigma}}$, the applied stress, affected by the damage and the initial state:

$$\begin{cases} \vdots \\ \vdots \end{cases} \quad \text{with} \quad \begin{cases} \vdots \\ \vdots \end{cases}$$

The differentiability of such an energy has been proved in [7]. The damage kinematics has not yet been chosen, but will be totally defined by the damage evolution law, in the same manner as the damage kinetics. Thus, the damage kinematics is driven by the following tensorial damage force:

$$\underline{\underline{Z_C}} = \underline{\underline{C}} \cdot \underline{\underline{Y_C}} \cdot \underline{\underline{C}} \quad \text{with} \quad \underline{\underline{Y_C}} = \frac{\partial e_d}{\partial \underline{\underline{C}}} = \underline{\underline{\sigma}}^+ - \underline{\underline{\sigma}}^-$$

And the damage kinetics by the following scalar damage force:

The damage evolution law is now defined by:

with

Where the parameter **a** drives the damage anisotropy, the parameter **n** being chosen large enough; and where , $\check{z}_0$ and $\check{z}_1$ are the parameters that must be identified on experiments. This model has been implemented into the LMT C++ platform [16, 17] (see Figure 2). Work is in progress to introduce a delay effect for the damage evolution, in order to control its localization.

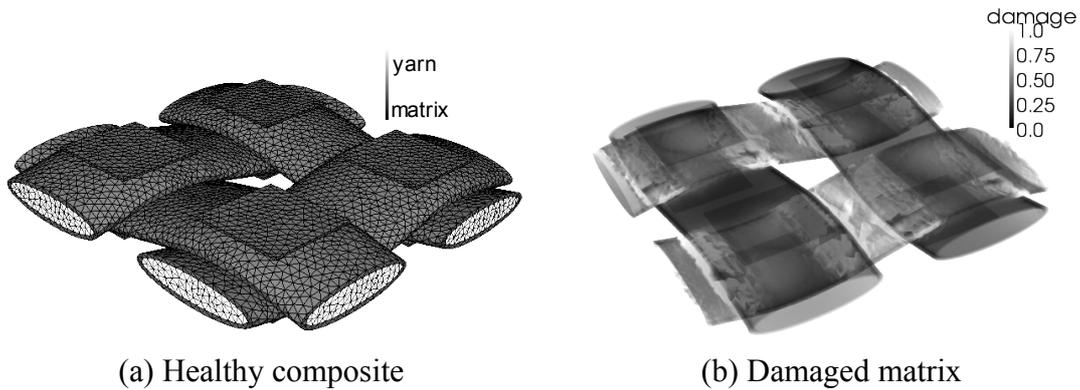

(a) Healthy composite              (b) Damaged matrix

Figure 2: Meso-cells at yarn scale ([9]'s meshes)

**Yarn cracking**

The fundamental points concerning the transversal intra-yarn matrix cracks are the following: i) they are orthogonal to fibers; ii) they do not break the fibers, because of the [C] interphase [2, 3], so there is still a stiffness in the material at the location of a crack; iii) the behaviour of the crack (opening, inelastic deformation, etc) is driven by the fiber/matrix debonding process [3, 8, 13].

In order to derive a macroscopic model from a microscopic representation of these cracks and associated debonding zones, a micro-macro bridge [14] is build. The theory is described here in the unidirectional case, and stress fields are calculated by shear lag analysis with a constant shear stress [13], written .

Let us consider a fiber-matrix system under traction, with multiple cracked matrix and associated debonding zones (see Figure 3). Following [18]'s approach, the internal energy of such a system is separated between the stored and the recoverable elastic energies:

$$E^i = E^s + E^r$$

This separation of energy is made for any given state (**A**) by separating the problem defining the state (**A**) into a purely frictional problem defining the conceptual state (**A**⁰) and a purely elastic problem defining the linear path (**AA**⁰) (see Figure 3). The frictional problem has inelastic deformation, written $\bar{\varepsilon}^i$, internal variable of the homogenized model, and auto-balanced stresses defining a stored elastic energy. Moreover, in the hypothesis of the shear lag the elastic energy of the $(AA^0)$ problem simply writes:

Where $\overset{\rightharpoonup}{E}$ is the average Young modulus of the composite system. Finally, the behaviour law of the composite derives from its recoverable elastic energy potential:

A major difficulty is that the inelastic deformation $\varepsilon^i$ does not derive from the stored energy potential, because it is not a state potential but indeed depends on the load history. Then, additional internal variables must be introduced. However, for most practical cases it is possible to derive $\varepsilon^i$ with the use of very few additional internal variables. To do so, let us write:

$$\varepsilon^i = \frac{u^i}{l}$$

Where $u^i$ is the inelastic displacement associated to a single crack, and $l$ is the average crack spacing. Thus, calculating $\varepsilon^i$ relates to calculate $u^i$ and $l$.

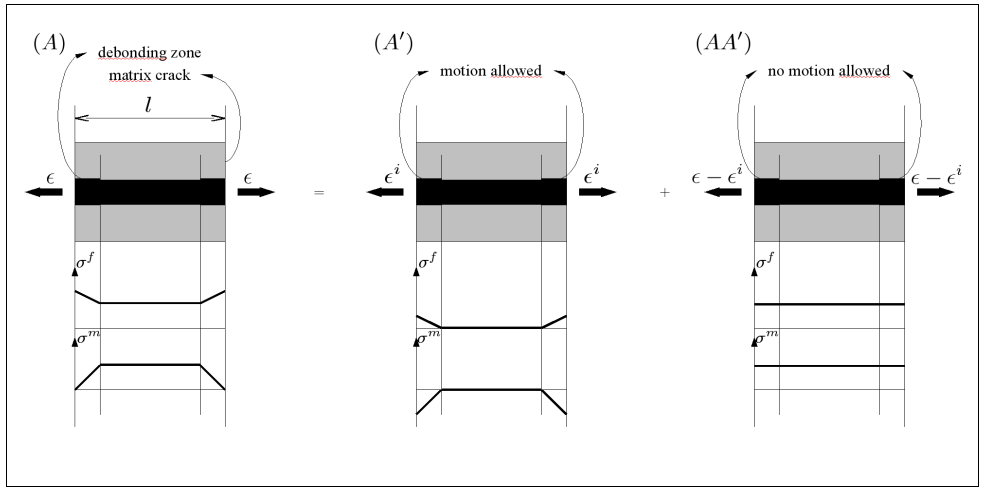

Figure 3(a): Decomposition of the frictional problem $(A)$ into a purely frictional problem $(A^0)$ and a purely elastic problem $(AA^0)$, stress fields

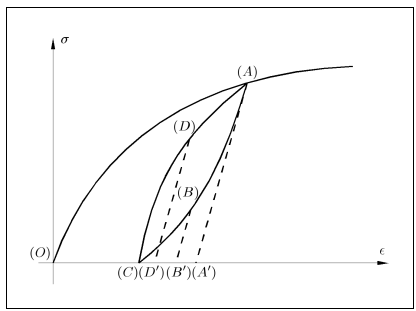  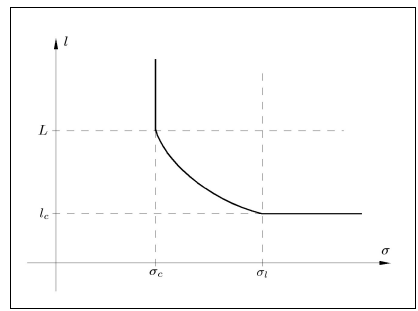

Figure 3(b): Stress-strain curve        Figure 4: Crack spacing law

First, the shear lag approximation gives simple expressions for $\underline{u}^i$ in most practical cases. For instance (see Figure 3), along the (OA) path, $\underline{u}^i = \dfrac{r_f\, E_m^2\, V_m^2\, \sigma^{3/4}}{\bar{E}^2 E_f\, V_f^2\, \tau}$; along the (AC) path, $\vdots$ ; and along the (CA) path, $\underline{u}^i = \dfrac{r_f\, E_m^2\, V_m^2\, \sigma^{3/4}}{2\bar{E}^2 E_f\, V_f^2\, \tau}$.

It is of importance to notice that in such complete unloading-reloading paths, only one additional internal variable is required to describe the stress fields: the maximum applied stress, written $\sigma^A$. For more perturbed loads, additional care must be taken.

Second, the average crack spacing can be roughly approximated by the following law (see Figure 4):

$$\vdots \quad \text{with} \quad \vdots$$

Where $L_0$, and $m$ are classical Weibull coefficients, and $L$ the length of the composite tow.

Finally, the microscopic model is written in a completely macroscopic formalism, which means that it can be simulated as any continuum model. However, it worth pointing out that this model is only based on microscopic known parameters and that its internal variables are strongly linked to the microscopic representation of the cracks. For instance, at each step of the calculation, the density and opening of the cracks are perfectly known. The Figure 5 presents the result of the model with [19]'s parameters, and validate the approach. Work is in progress to extend it to the tridimensional case, and to implement the whole model into the LMT C++ platform.

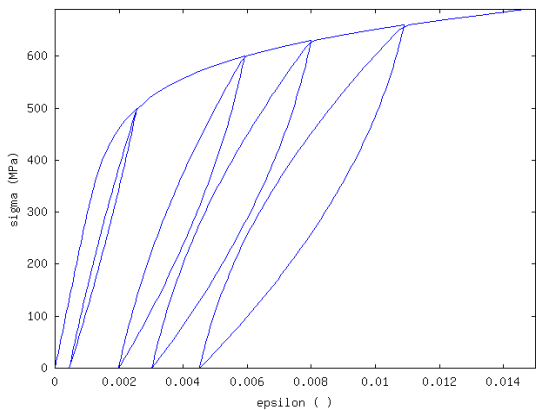

Figure 5(a): Stress-strain curve with unloading-reloading

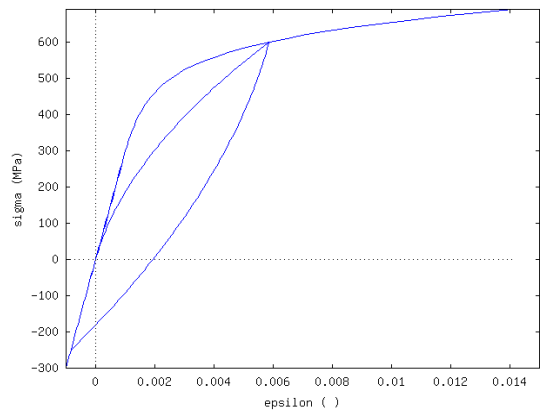

Figure 5(b): Stress-strain curve with traction-compression

# CHEMICAL PART OF THE MODEL

SiC fibers used in CMCs suffer from sub-critical cracking –they break under their strength– under oxidizing atmosphere [4]. This is due to surface defects sub-critical propagation –the classical critical stress intensity factor is not reached, but the crack still propagates– (see Figure 7a). This propagation is induced by a degradation of the crack tip through a stress-assisted oxidation reaction [4, 20]. A modeling framework is presented to simulate sub-critical propagation of cracks, as a replacement of the widely used one based on a Paris-like law [4, 20]. It belongs to fracture mechanics and is based on a pure material law describing the degradation of the material resulting from oxidation. It is presented in detail in a forthcoming paper.

Let us consider a section of the fiber with a surface defect (see Figure 7b). The size of the defect is written $a$. This defect induces a stress singularity in the fiber, which stress intensity factor is:

$$K = \sigma \sqrt{\pi a} \cdot Y$$

Where $\sigma$ is the applied stress on fiber and $Y$ the shape factor [4, 20]. This law gives the value of the initial defect size $a_0$ and the critical defect size $a_c$ in function of the probabilistic strength of the fiber $\sigma_R$ and the applied stress $\sigma$ (see Figure 7b):

$$a_0 = \frac{1}{\pi} \left( \frac{K_c}{\sigma_R Y} \right)^2$$

$$a_c = \frac{1}{\pi} \left( \frac{K_c}{\sigma Y} \right)^2$$

Where $K_c$ is the classical critical stress intensity factor (without oxidation).

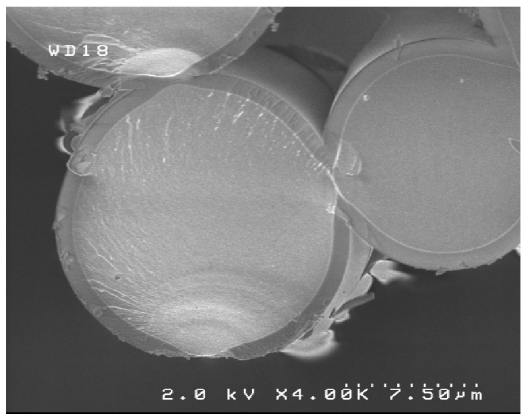

(a) Observation [22]: Surface defect and oxide layer

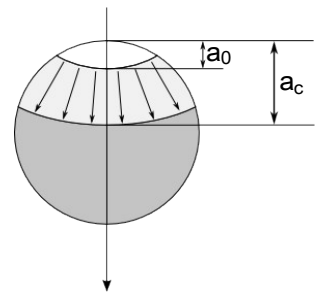

(b) Scheme: surface defect growth

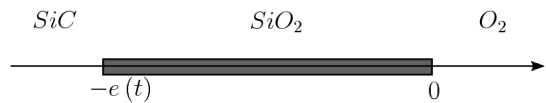

(c) Scheme: oxide layer growth

Figure 7: Fiber section with surface defect and oxide layer

The sub-critical propagation results from the equilibrium between two mechanisms: i) the oxygen arriving on the crack tip, which reduces the local mechanical properties of the material, which favors the propagation; ii) the defect growth itself, which brings an healthy crack tip, which penalizes the propagation.

This equilibrium is well described by a pure material law that gives the evolution of the sub-critical stress intensity factor –smaller than the critical stress intensity factor because of the oxidation, written $K_{sc}$– with regards to the two parameters driving the intensity of each mechanism, say the defect velocity –written $v$– and the oxygen flux arriving on the fiber –written $Å$–:

Where $¸$ and $n$ are model parameters that must be identified. Moreover, the crack propagates within the framework of fracture mechanics:

$$K = K_{sc}$$

Taking a constant oxygen flux along time, this framework falls on the classical Paris-type law used in [4, 20]. However, in this study the oxygen flux is given by an adaptation of Deal-Grove model [21] to the $SiC_{(s)} + O_{2(g)}$ ! $SiO_{2(s)} + CO_{2(g)}$ reaction (see Figure 7c). This diffusion/reaction problem predicts that the growth of the oxide layer around the fiber is first linear and driven by the reaction at the fiber surface –oxygen flux is then constant and linked to the reaction coefficient–, and then square rooted and driven by the diffusion in the oxide layer –oxygen flux is then decreasing and linked to diffusion coefficient and thickness of the oxide layer–. Consequently, it will allow the model to deal with reaction- as well as diffusion-controlled propagation – which increases fiber lifetime at high temperature–, without the addition of parameters as it is done by others [22]. Moreover, the model can naturally deal with variable fiber environment –temperature, oxygen concentration in surrounding air–, which is necessary to introduce self-healing mechanisms.

The predictions of the model are presented Figure 8, and compared to [4]'s experiments. It is worth pointing out that the actual model is very close to experiments, even during diffusion-controlled propagation at high temperature, with only two parameters.

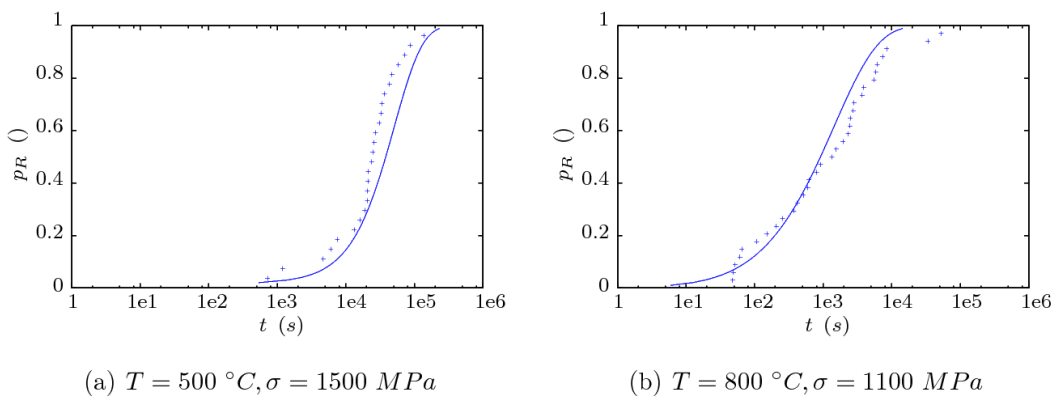

(a) $T = 500\ °C, \sigma = 1500\ MPa$  (b) $T = 800\ °C, \sigma = 1100\ MPa$

Figure 8: Fibers rupture probability versus time for different loads, model vs. experiments [4]

# CONCLUSION

A first version of a virtual material for lifetime prediction on self-healing ceramic matrix composites has been presented. The model is divided into two main parts: a mechanical one and a chemical one.

The mechanical part provides the key input data for the chemical part: stress on fibers and crack network morphology. The analysis is mainly done at yarn scale:

- The representation of the material's morphology directly uses [9]'s work.
- Inter-yarn matrix cracks are handled through the use of an anisotropic continuum damage model. Work is in progress to control the damage localization.
- Longitudinal intra-yarn matrix cracks will appear explicitly through the use of potential crack surfaces. This pas has not been achieved yet.
- A micro-macro bridge is build to handle with transversal intra-yarn matrix cracks continuously at yarn scale, keeping information on explicit micro cracks. The unidirectional case has been validated, and work is in progress to extend the bridge to the tridimensional case.

The chemical part provides lifetime prediction for the material. The analysis is mainly done at fiber scale: a fiber lifetime model coupling mechanical load and chemical environment is built in a powerful framework for crack sub-critical propagation. It has been validated, and work is in progress to handle with the complete self healing process.

# ACKNOWLEDGEMENTS

This work has been partially funded by French Army, through a grant accorded to M. Genet, and by Snecma Propulsion Solide.